# *DREAMTIME ASTRONOMY*: DEVELOPMENT OF A NEW INDIGENOUS PROGRAM AT SYDNEY OBSERVATORY

Geoffrey Wyatt[1], Toner Stevenson[1], and Duane W. Hamacher[2]

[1]Sydney Observatory, Watson Road, Observatory Hill, Sydney, NSW, 2000, Australia
[2] Nura Gili Indigenous Programs Unit, University of New South Wales, Sydney, NSW, 2052, Australia
Corresponding Email: geoffw@phm.gov.au

**Abstract**

The Australian National Curriculum promotes Indigenous culture in school education programs. To foster a broader appreciation of cultural astronomy, to utilise the unique astronomical heritage of the site, and to develop an educational program within the framework of the National Curriculum, Sydney Observatory launched *Dreamtime Astronomy* – a program incorporating Australian Indigenous culture, astronomy, and Sydney's astronomical history and heritage. This paper reviews the development and implementation of this program and discusses modifications following an evaluation by schools.

## 1 INTRODUCTION & BACKGROUND

Sydney Observatory is the oldest surviving observatory in Australia, constructed at various stages starting in 1858. In 1982 Sydney Observatory ceased its research program and was incorporated into the Museum of Applied Arts and Sciences (MAAS) (Kerr, 1991). Whilst education has long been a role of the Observatory, science education and the heritage of astronomy are now the primary focus. As a public observatory and museum, Sydney Observatory offers unique and site-specific educational programs in astronomy, meteorology, archaeology, and most recently on the astronomical knowledge of Aboriginal Australians, a topic dubbed "Aboriginal Astronomy."

Since 1997 Aboriginal astronomy has been part of the exhibition and all education tours have included some Aboriginal content, mainly in the planetarium component. Within the domed, darkened room, stars are digitally projected to create a virtual night sky. Aboriginal constellations, such as the Emu in the Sky, are projected on the dome and the astronomy guides recount stories from communities such as Yolngu (Northern Territory) and Murri (Queensland).

Over the past decade there has been a significant increase research about the extent by which Indigenous Australians developed astronomical knowledge (e.g. Cairns and Harney, 2004; Fredrick, 2008; Hamacher, 2012; Norris and Norris, 2008). This has coincided with changes in the school curriculum and a desire to increase the number of student visitors to Sydney Observatory. Day school visitation increased nearly 45% to 16,000 between 1998 and 2010, but has plateaued since 2010 (Figure 1). Student numbers for night programs have remained static at around 2,000 per annum.

A decision was made to embrace the new research in Aboriginal astronomy and develop an education program around the subject. This goal was to increase education visits and further enhance the appreciation of Indigenous Australians as the world's oldest astronomers (Norris, 2008).

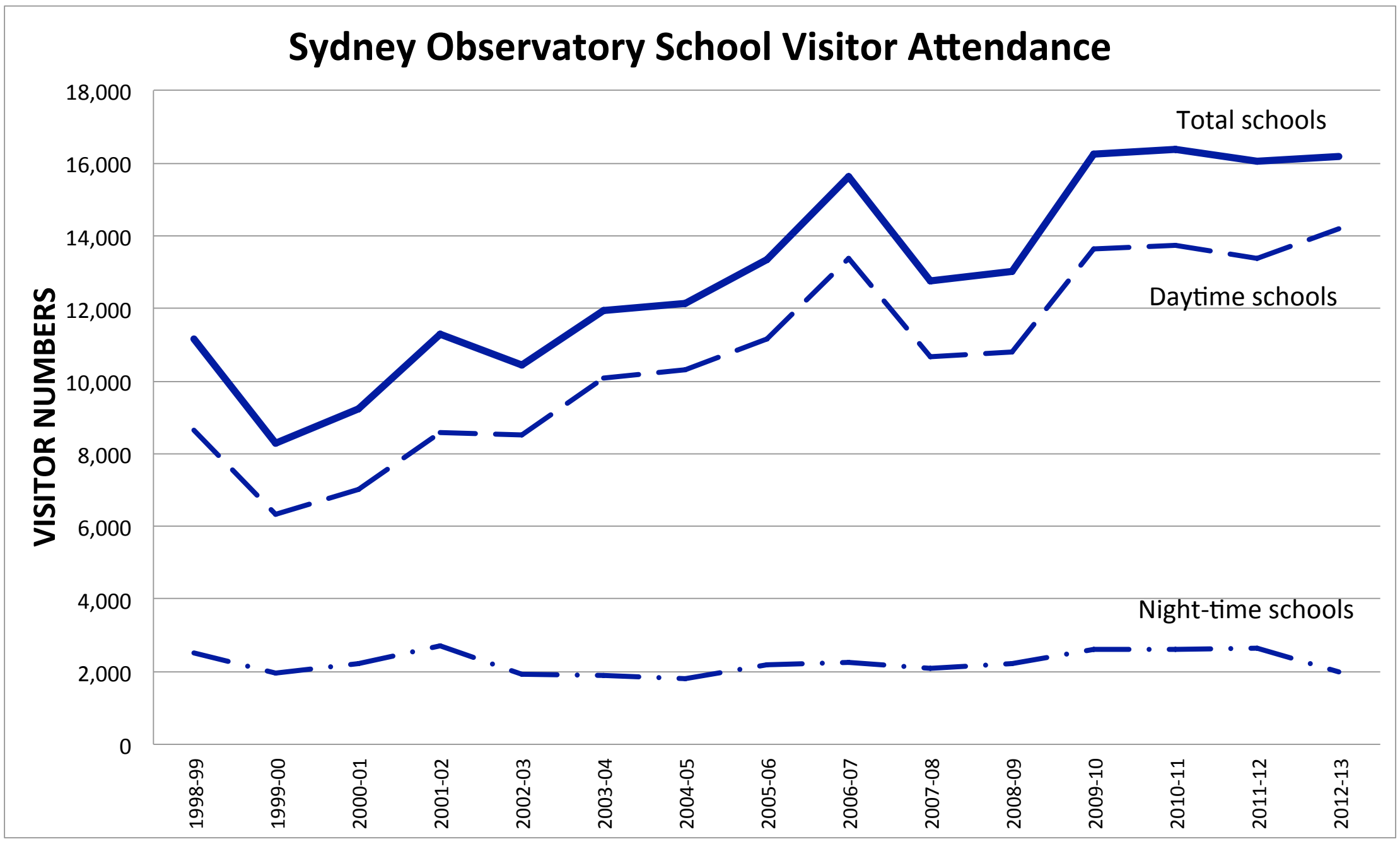

*Figure 1: Education visits to Sydney Observatory between 1998 and 2013*

## 2 PREVIOUS ABORIGINAL ASTRONOMY PROGRAMS

In 1997 a new exhibition marked the start of the Observatory's presentation of Indigenous astronomy. The small theatre exhibit, named *Cadi Eora Birrung* (meaning "*Beneath the Southern Sky"* in the language of Cadigal people of Sydney Cove), includes an immersive space where videos of astronomy-themed Dreaming stories are projected using a 'fishers-ghost' technique. The animations were developed by Aboriginal artists and recorded by Aboriginal actors[1]. Nearby, an interactive computer display allows the user to choose between Aboriginal and Western constellations visible during different seasons of the year[2].

In 2012, we developed an exhibit that featured meteorites from New South Wales, the Northern Territory, and Western Australia. The exhibit[3] includes both scientific and Aboriginal views of the Henbury crater field (NT) and Wolfe Creek crater (WA). The display also features a painting of the Wolfe Creek crater (called *Kandimalal* in the Jaru language) by a noted Jaru artist. We have received positive feedback about the display from visitors.

The installation of extensive Western heritage and Aboriginal exhibits in 1997 marked the beginning of a trend of increasing visitor attendance to Sydney Observatory (Figure 2). This increase has, on the most part, been steadily continuing ever since. There is no evidence to suggest one part of the Sydney Observatory experience is more visited than another. As it is a small building, visitors tend to see a majority of the exhibits. The *Cadi Eora Birrung* exhibit is referenced in tourism guides (Atkinson, 2012) and on the City of Sydney's self-guided walking tour and website (Aboriginal Cultural Attractions, 2014). On average, tourists account for 42% of the visitors to Sydney Observatory.

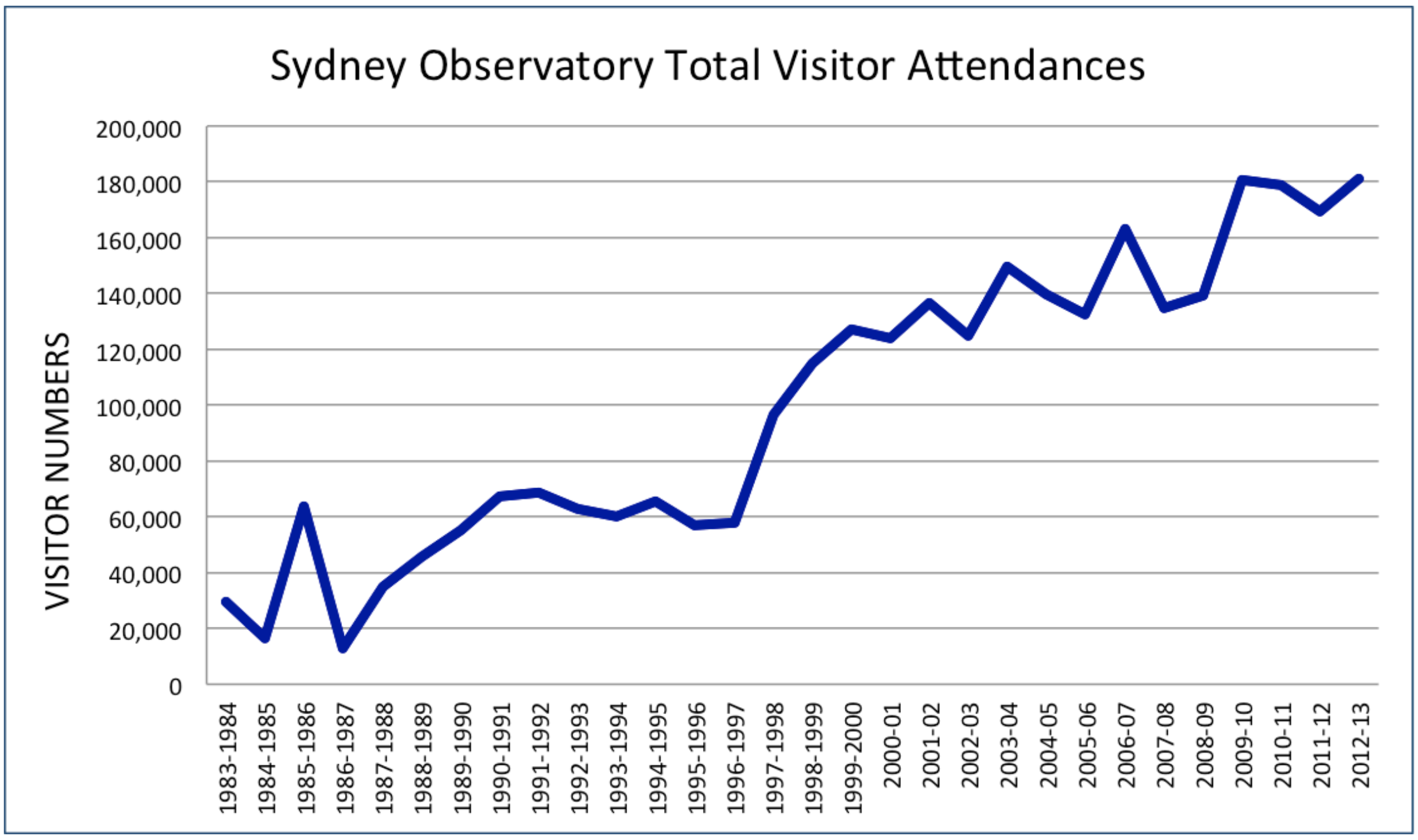

*Figure 2: Thirty years of visitor attendances at Sydney Observatory showing the impact of the 'By the light of the Southern Stars' exhibition and 'Cadi Eora Birrung'.*

## 3 PROGRAM DEVELOPMENT

The development of the new Australian Curriculum (ACARA, 2013), guided by the Melbourne Declaration on Educational Goals for Young Australians in 2008[4], stressed the importance of a cross-curriculum priority on Indigenous histories and culture. This is the primary motivation for developing an Aboriginal astronomy program, along with educating the general public about the complexity of Indigenous astronomical knowledge.

The *Interactive Experience Model* (Falk and Dierking, 1992), where the personal, social, and physical intersect, is a dynamic way for museums to achieve successful learning experiences for school children using their unique attributes outside the school environment. Sydney Observatory has adopted this hands-on approach as central to all of its education programs (Lomb and Stevenson, 2008). With an Indigenous astronomy program, our aim is to achieve what Canadian astronomy educator, John Percy, calls 'minds-on' education.

Percy (2008: 20) considers methods used by ancient cultures as an ideal way of engaging students in the 'big picture' concepts of astronomy:

> "*The First people to 'do astronomy' were the pre-technological cultures who were able to determine time, date, and direction from the sky. They made and recorded observations, archived and used them… one could argue that these simple observations are the most relevant kinds for most students to make.*"

This way of thinking made it possible to tie the cultural aspects of Aboriginal astronomy and the scientific approach of astrophysics and cosmology together with a common thread of observation, recording, and analysis.

Originally titled *Shared Sky*, and later renamed *Dreamtime Astronomy,* the program highlights the diverse astronomies of Aboriginal people and explores the ways in which they used the stars for navigation, seasonal calendars, food economics, ceremony, and social structure. The framework of Aboriginal cultures, including their beliefs, spirituality, laws, and oral traditions, is called the *Dreaming,* or *Dreamtime*. It can be viewed as a period in the distant past when ancestors created the land, sky, animals, and plants. It can also represent a present or future reality in which people interact with ancestor spirits (Isaacs, 1980). The *Dreamtime* contains information regarding the daily practices, social structure, and knowledge of the community. The *Dreamtime* is synonymous with Traditional Knowledge but is unique to Australian Aboriginal cultures. This is the reasoning behind naming the program *Dreamtime Astronomy*.

A majority of the Aboriginal content for this program is based upon the ethnographic work of William E. Stanbridge, who published a paper on the astronomical traditions he learnt from the Boorong clan of the Wergaia language group in northwestern Victoria (Stanbridge, 1861). This was chosen because it represents the earliest comprehensive description of Aboriginal astronomy and is one of the most studied Aboriginal views of the night sky (Hamacher and Frew, 2010; Morieson, 1996).

To implement the program, we needed to determine how to make Aboriginal content relevant and memorable to students who have typically had no Aboriginal astronomy exposure prior to the visit to Sydney Observatory. We designed three activities to accomplish this.

## 4 PROGRAM IMPLIMENTATION

The *Dreamtime Astronomy* program, like most tours at the Observatory, lasts for 1.5 hours and involves three activities, taking 30 minutes each.

### 4.1 ACTIVITY 1: USING STELLARIUM

In this activity, students use modern technologies familiar to them, including computers, tablets, and phones, to use interactive astronomy apps and software packages, such as *Stellarium*[5]. Stellarium allows the user to simulate the sky on their device. This is particularly important as students can set the view to the night of their visit to the Observatory.

The positions of the planets and Moon along the ecliptic are used to highlight students' existing 'latent' knowledge of constellations, which is typically limited to those of the Zodiac. The software allows students to learn to use the stars to tell time and find direction. This will provide them with a foundational understanding of how the sky works and give them a chance to play with the software, for example seeing what objects were in the sky the night of their birth.

## 4.2 ACTIVITY 2: MAKING A PLANISPHERE

The second activity involves students making a planisphere[6] (Figure 3) that they can take home. The planisphere includes two discs: the first shows Western names of celestial objects. It is used to help the students determine what stars are in the sky throughout the night at various times of the year, find directions, and tell time. Having explored the student's knowledge of constellations related to Greek mythology, the students replace the inner star-wheel with the second disc. This disc shows Boorong names of celestial objects, accompanied by a small pamphlet with relevant information about Boorong astronomy. It is red, black, and yellow to reflect the colours of the Aboriginal flag.

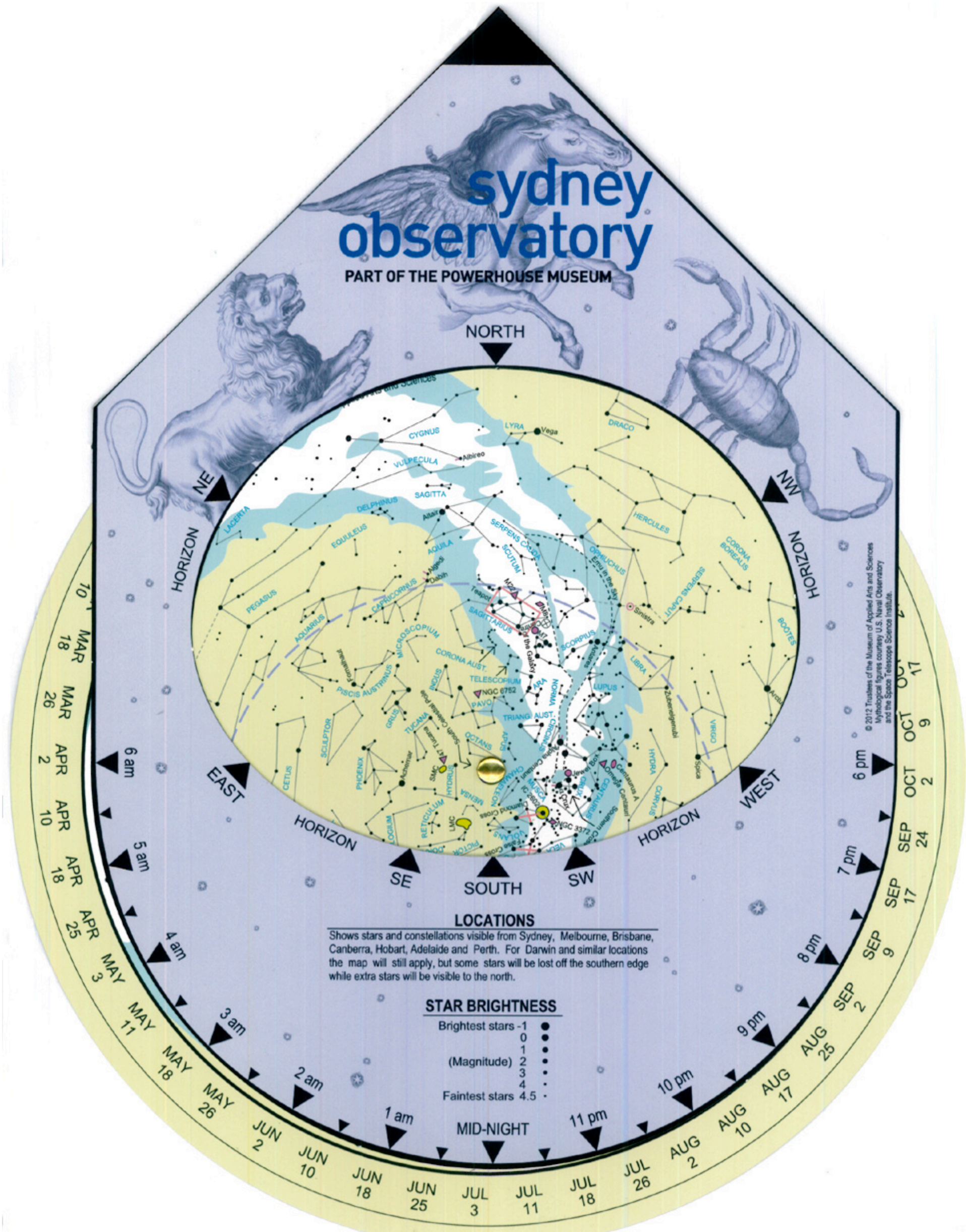

*Figure 3: A Western planisphere, developed by Dr Martin Anderson.*

With the aid of an Indigenous language map of Australia, students are made aware of the diverse range of Aboriginal cultures and astronomies across the continent and emphasise the lack of a single Aboriginal "astronomy" or viewpoint.

Faint, half-tone representations of the Southern Cross, Scorpius, and Orion are included on the Boorong star-wheel as a reference (Figure 4). These familiar constellations permit students to draw observational correlations between what they already know and what they are learning. Boorong stories that go with the stars are recited and projected on an interactive whiteboard. Problem solving activities are presented to the students, such as identifying which Boorong star is visible at particular times of year and its significance to the people. Examples include the appearance of Arcturus (*Marpeankurrk*) high in the northern sky at dusk when the Boorong people would collect wood-ant larvae, or the rising of Vega (*Neilloan*) at dusk when the mallee-fowl birds began building their nests (Stanbridge, 1861). The accompanying pamphlet provides these details, explaining what these stars meant to the Boorong people.

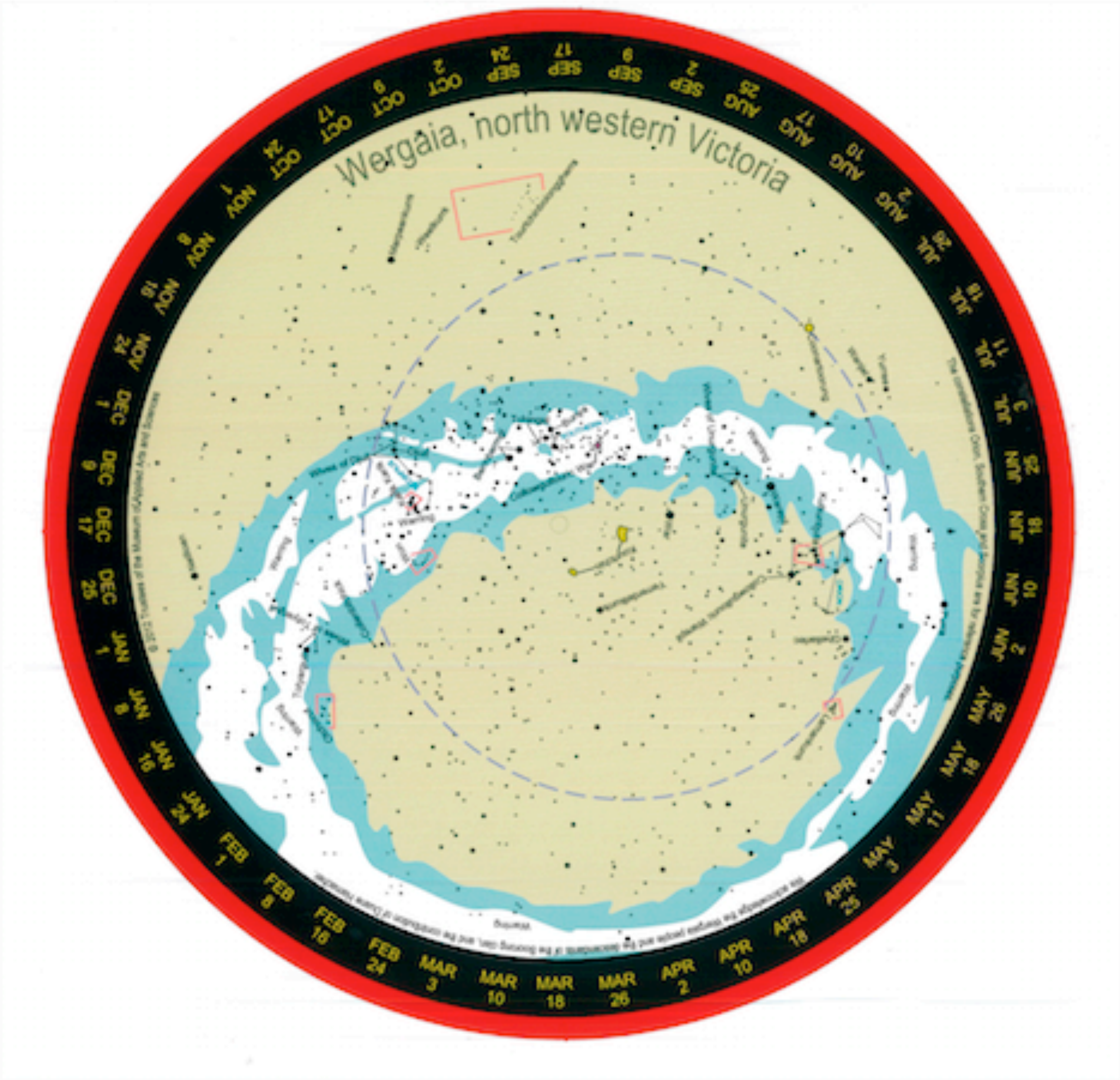

*Figure 4: Boorong planisphere inner-wheel by Dr Martin Anderson.*

## 4.3 ACTIVITY 3: MAKING A SIGNAL STICK

The third component of the program addressed forms of non-verbal communication used by Aboriginal communities that also highlighted the Observatory's role in the young colony. Message sticks were used by Aboriginal people to send information to communities great distances away (Howitt, 1904). This information might include the time and place for a corroboree (ceremony), with the former denoted by lunar phases (Hamacher and Norris, 2011).

The Observatory played an important role in signalling from about 1814 onwards, although this faded in modern times. A newly erected flag mast[7] at the Observatory now flies flags that not only replicate old signals, but also flags that communicate Sydney's expected maximum temperature, phase of the moon, constellations visible, planets visible, and special astronomical events such as meteor showers, comets, equinoxes, and solstices. In the words of the team led by George Oxenbridge that made the flag mast, it became a "*…giant message stick for Sydney*," (Figure 5).

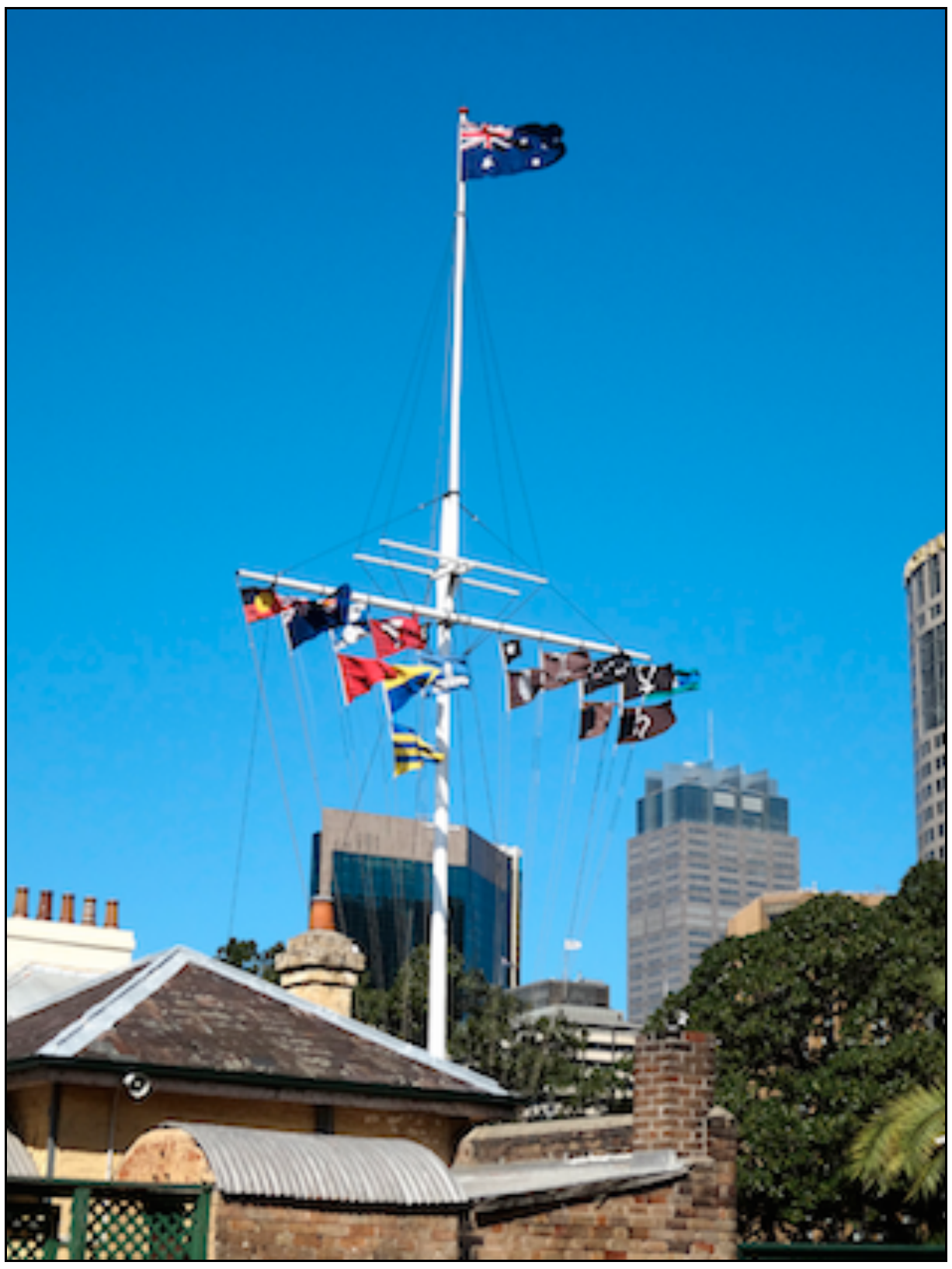

*Figure 5: The 2008 flag mast recreation by the Bruce and Joy Reid Foundation. Photo: Geoff Wyatt*

It was this idea about signals and message sticks that was used to develop the third (though not entirely astronomical) component of the program. After consultation with the Australian Museum in 2011, the more neutral term 'signal stick' was used, as message sticks sometimes contained information that was gender sensitive (e.g. "Men's business" or "Women's business", such as initiation ceremonies). A brief history of signal sticks would be presented with a two-fold intention:

1) To show their analogous connection to modern passports, and
2) That they represent a complex form of non-verbal communication, comparable to Fort Phillip's flag mast.

To inspire the students and serve as a visual prompt, a sample of symbols from various cultures was projected during this section of the tour. Symbols were borrowed and included some of Aboriginal origin[8], others from Asian and Middle Eastern cultures, along with commonly used symbols such as stars, the heart, male and female symbols, which astronomically represent Mars and Venus respectively (Figure 6).

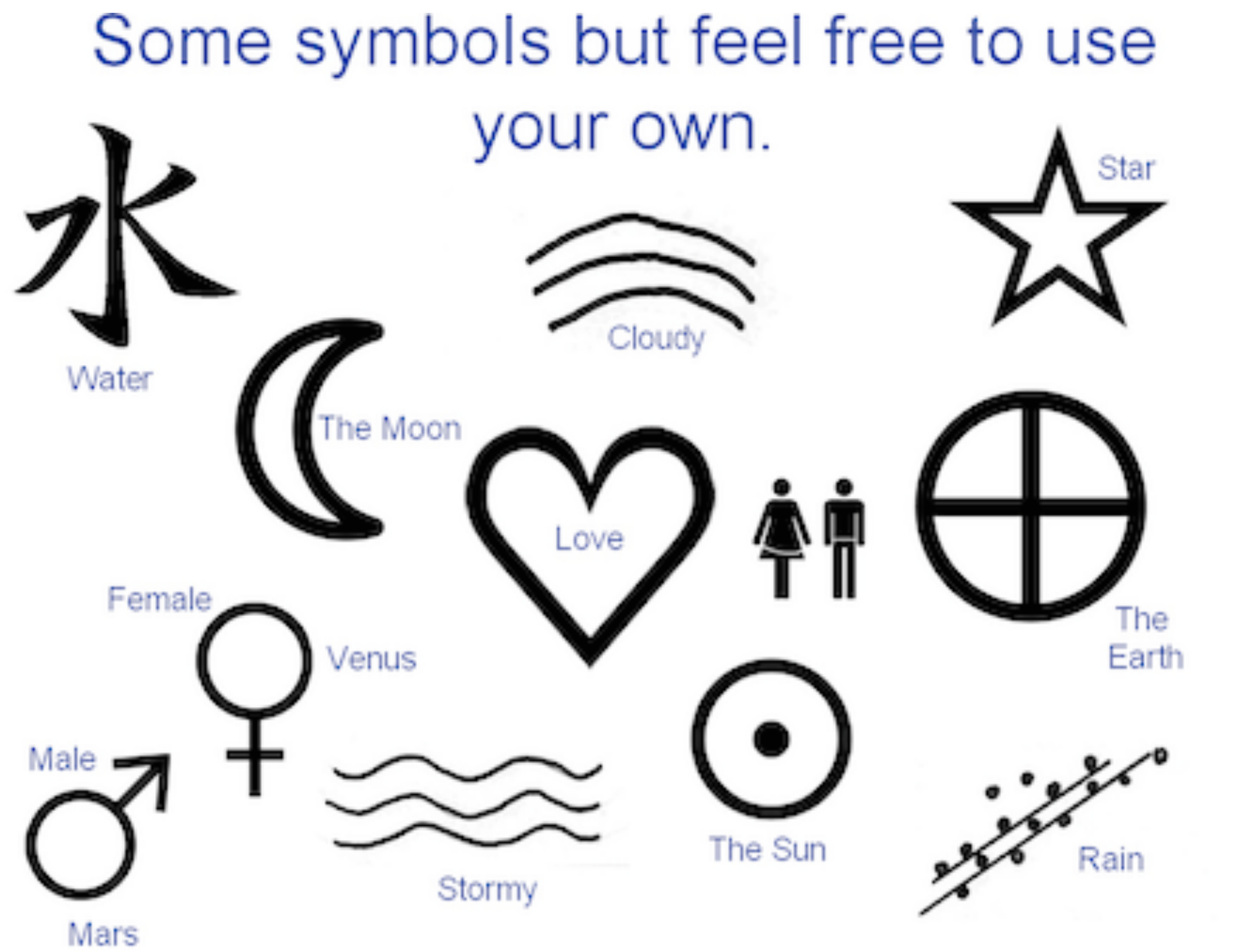

*Figure 6: Sample of multi-cultural symbols. Image: Geoff Wyatt*

Students were then given the opportunity to create a signal stick to take home (Figure 7). Each student could decorate their signal stick with symbols that would mean something to themselves or immediate circle of friends and family. It would not need to be intelligible to those other than the intended recipient and could be of any design.

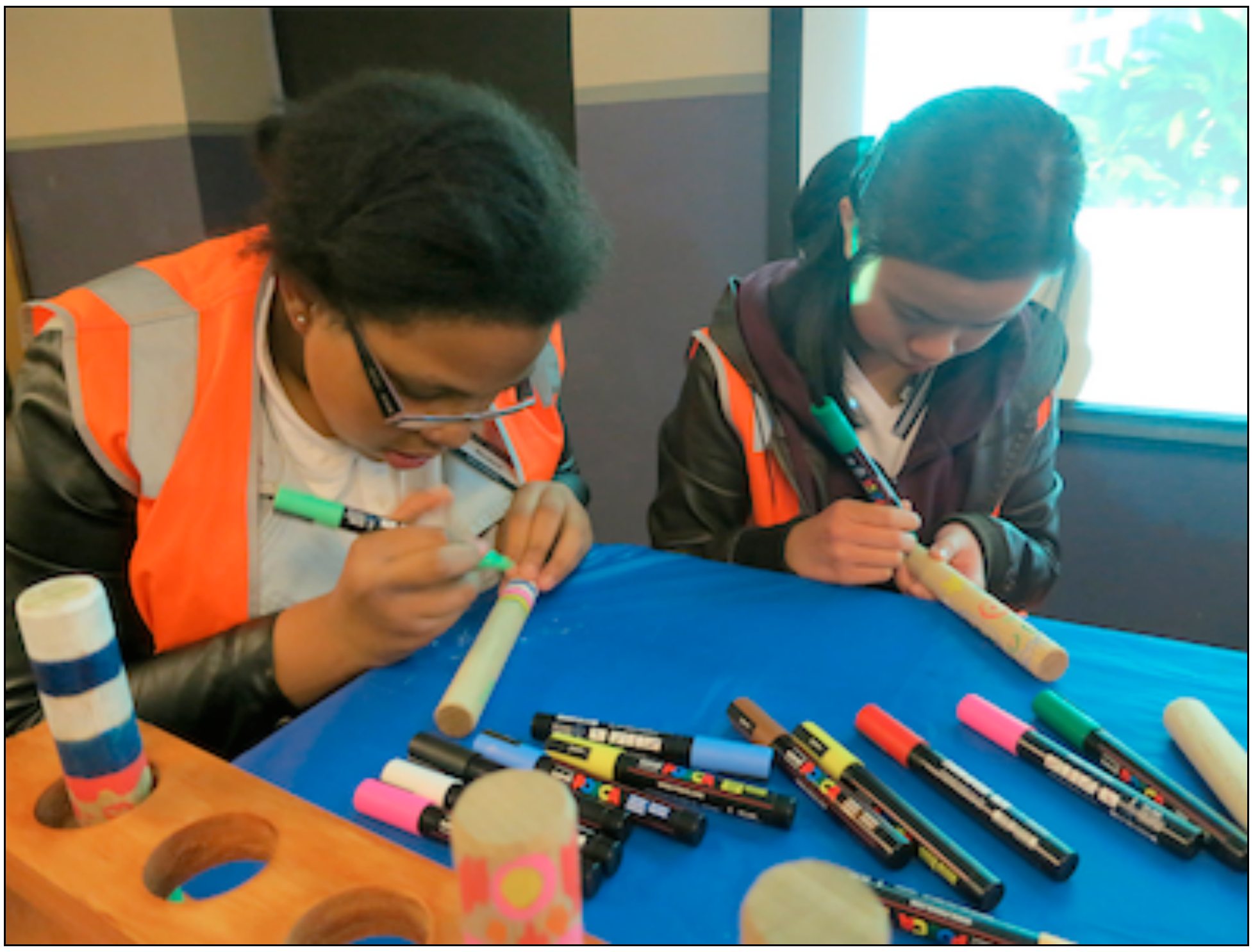

*Figure 7: Sample Signal sticks. Photo: Geoff Wyatt*

## 5 PROGRAM EVALUATION

In 2013 the program was tested with a sample of four schools with students ranging from Year 2 to Year 4 (aged 7 to 10). Teachers were asked to provide feedback and the feedback was predominantly positive. The use of modern technologies, such as computers, tablets, and smart phones, was well received and the students enjoyed learning with Stellarium.

There were a few major points of criticism from the teachers, students, and guides:

1) The planisphere could be improved by using local Aboriginal astronomy content (Sydney region) rather than content from western Victoria or other parts of Australia.

2) Students were disappointed that they did not get to use the planetarium.

3) The signal sticks were an enjoyable activity, but were the least relevant to an experience at Sydney Observatory.

4) The students were extremely disappointed that they did not get to look through a telescope.

5) Indigenous guides delivering the content would improve the program.

This feedback enabled us to modify and improve *Dreamtime Astronomy*.

## 6 PROGRAM MODIFICATIONS

### 6.1 MODIFIED ACTIVITY 1

Students and teachers found Activity 1 (using Stellarium) to be educational, enjoyable, and engaging. But we decided to combine it with Activity 2 (making a planisphere), as they complimented each other. We also decided to replace the Boorong astronomy content in the planispheres with Sydney Aboriginal astronomy content. Unfortunately, little has been published on Aboriginal astronomy in the Sydney region, so we developed a partnership between the Observatory and the Nura Gili Indigenous Programs Unit at the University of New South Wales to address this issue.

The UNSW partnership involves a co-author of this paper (Hamacher), who is an academic at Nura Gili and works casually at the Observatory as an astronomy educator. Hamacher teaches an undergraduate unit called *ATSI 3006: Astronomy of Indigenous Australians*. A major assessment for students enrolled in the course is to research an area of Australian Indigenous astronomical knowledge and develop educational materials utilising that research. Students in *ATSI 3006* will research Aboriginal astronomical traditions in the Sydney region and develop a planisphere based on these findings. Preliminary findings have already revealed a number of Aboriginal stories of the sky from the Sydney region, including the Dharug, Dharawal, Gundungurra, and Kuringai peoples. This content will be incorporated into the *Dreamtime Astronomy* program in mid-2014.

We also realised the pamphlet containing Aboriginal astronomical information was not ideal for the students, as most were left behind. But the students valued the planispheres and kept these. When the students complete their planispheres, the guides show them how to use it. They then demonstrate the connection between the rising or setting times of stars and their connection to seasonal cycles using Stellarium. The students are given a short demonstration and shown how to download the program to their devices. The pamphlet was discontinued and instead the information will be presented on a webpage from mid-2014.

## 6.2 MODIFIED ACTIVITY 2

We determined that the low astronomical significance of the signal stick activity did not reflect an ideal approach to teaching the students about Aboriginal astronomy. Additionally, the sticks did not serve a useful purpose once the students left the Observatory. In fact, many were left behind. The students' biggest expectation was to look through the telescopes in the dome, and they were extremely disappointed that the program did not include this. This came as no surprise as Observatory staff are anecdotally aware that all visitors, regardless of the program, have a strong desire to look through the telescopes in the domes.

We replaced the signal stick activity with a telescope viewing in one of the domes. Daytime programs will involve students viewing the Sun through a solar telescope, enabling staff to draw attention to the astronomical significance of the solar symbol in the Aboriginal flag while allowing the students to observe solar flares, sunspots, and learn about the physical structure of the Sun.

For nighttime programs, students will view astronomical objects of significance to Aboriginal people. This may include the Moon, planets, stars and star clusters. It gives the guides a chance to discuss the Aboriginal and scientific significance of these objects. One aspect relates to the colour of stars. The colour of a star can tell astronomers about its surface temperature and age. In many Aboriginal cultures, colours of stars might signify an association with a particular animal or plant. For example, in the Western Desert the red star Antares in Scorpius was associated with the red-tailed black cockatoo. When Antares rises at dusk in May, it informs Aboriginal people that the first clutches of eggs will begin hatching (Hamacher and Leaman, 2014).

## 6.3 MODIFIED ACTIVITY 3

In response to the student's disappointment that they did not get to use the planetarium, we developed a new activity that incorporates use of the planetarium and emphasises some of the relevant local history.

The earliest information we have about Aboriginal astronomy was recorded by Lieutenant William Dawes, the astronomer on the First Fleet that arrived in Sydney in 1788. He founded the first observatory in Australia one the shores of the harbour at the place now called Dawes Point, which lies at the southern base of the Sydney Harbour Bridge. During the short time he resided in the colony, he befriended a young Aboriginal woman named Patyegarang who taught him the local language and customs. He kept detailed journals, which contain local Dharug names of the Sun, Moon, Milky Way, and Magellanic Clouds (Dawes, 1788-1791).

The new activity involves giving each student a worksheet that incorporates Sydney Aboriginal astronomy and the interactions between Dawes and Patyegarang. Aboriginal names of astronomical objects are written on the worksheet for the students to learn and use. They are then taken into the planetarium, where they experience Sydney's twenty first century light polluted sky. As the guides share Aboriginal stories of the sky, the planetarium is set to 1788 conditions enabling the Milky Way to be seen clearly as Dawes and Patyegarang would have seen it. This gives the students an opportunity to learn by experience, just like Dawes. Additionally the simulated night sky on the planetarium dome includes other visual queues, including images of animals, plants, and Aboriginal ancestors important to the story.

## 6.4 OTHER IMPLEMENTATIONS & FEEDBACK

It was deemed important for the Observatory to have the program delivered by Indigenous guides for the students to get a more appropriate experience and to promote Indigenous involvement in the program. In January 2014, the Observatory hired Indigenous guides to present the *Dreamtime Astronomy* program[8].

Australia contains two different Indigenous peoples: Aboriginal people and the Melanesian Torres Strait Islanders. Our current programs only focus on Aboriginal astronomy. Little has been researched on Islander astronomy in-depth, but related research shows that Islander culture contains a significant degree of astronomical knowledge (e.g. Sharp, 2003). A major grant to study Torres Strait Islander astronomy was awarded by the Australian Research Council (Hamacher, 2014). Information about Islander astronomy will be incorporated into the *Dreamtime Astronomy* program in the future to present a more comprehensive program in Indigenous astronomy as opposed to strictly focusing on Aboriginal astronomy.

The modified program, which will not include Sydney Aboriginal content until mid-2014, has been trialled for evening public tours. Feedback in the form of ratings on a social media website (Trip Advisor, 01 April 2014[9]) indicate its success:

> "*Got there early and toured museum. Very well done. Willy gave us stories from his people about the stars. Interesting, educational and fun. Recommend it.*"

Since developing an Aboriginal astronomy program at Sydney Observatory, the Observatory was listed on the 'Portal to the Heritage of Astronomy' website developed by UNESCO and International Astronomical Union as part of the Astronomy and World Heritage Initiative (Cotte and Ruggles, 2010). This listing provides further impetus to project astronomy as a science with cultural and heritage implications.

## 7 SUMMARY

Astronomy education at Sydney Observatory has enjoyed growth and strong visitation over many years whilst operating as an astronomical museum and adopting an Interactive Experience Model for its learning programs. New cross-curriculum opportunities have driven activity-rich experiences in Aboriginal astronomy. The *Dreamtime Astronomy* program will provide students and their teachers an opportunity to visit and engage the history and heritage of Sydney Observatory and Sydney Aboriginal astronomy. The ambition is to achieve a 'minds-on' experience for students, which engages them in astronomy through a cultural experience. Further evaluation will continue to inform the development and modifications of the program and its delivery.

For more than 150 years, Sydney Observatory has acted as a portal for European based science. It will now provide an opportunity for students to experience the astronomical knowledge and traditions of the oldest star watchers on the planet: Indigenous Australians.

## 8 ACKNOWLEDGEMENTS

We acknowledge the Cadigal people and pay our respects to elders past and present. This paper was originally presented by Geoffrey Wyatt at the 2013 Australian Space Sciences Conference held at the University of New South Wales in Sydney. We acknowledge the input and assistance of the following people for their work on the *Dreamtime Astronomy* program: Australian Museum education program personnel, Martin Anderson, Andrew Jacob, Nick Lomb, Katrina Sealy, and James Wilson-Miller. We also thank Dawn Casey for her support of the program development. Hamacher acknowledges Australian Research Council support (DE140101600).

## 9 NOTES

1. Aboriginal Nations created the Dreamtime cartoons, and the Powerhouse Museum especially filmed Aboriginal actors to introduce these in a 'fireside' setting. http://www.thedreamingstories.com.au/. The display was curated by Nick Lomb and James Wilson-Miller
2. The display was developed by Kathy La Fontaine
3. The exhibit was curated by Duane Hamacher and Katrina Sealy
4. Ministerial Council on Education Employment Training and Youth Affairs: http://www.mceecdya.edu.au/verve/_resources/national_declaration_on_the_educational_goals_for_young_australians.pdf
5. http://stellarium.org/
6. The planisphere was designed by Martin Anderson.
7. A re-creation of the flag masts was developed in 2008 through the generosity of the Bruce And Joy Reid Foundation
8. William "Willy" Stevens is the first Aboriginal guide to present *Dreamtime Astronomy*.
9. Pdm267, posted 1 April 2014, 'Dream makers tour', Trip Adviser, http://www.tripadvisor.com.au/ShowUserReviews-g255060-d256722-r199493592-Sydney_Observatory-Sydney_New_South_Wale.html#UR199493592

**ABOUT THE AUTHORS**

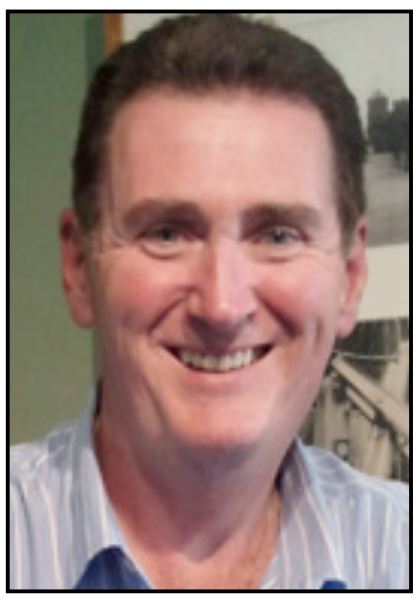

Geoffrey Wyatt is the Education Officer at Sydney Observatory, part of the Museum of Applied Arts & Science. He holds a Master of Education from the University of Sydney, and was awarded the 2004 Malcolm King award for staff development at the Powerhouse Museum and the NSW TAFE Medal for Conveyancing. He is an experienced astrophotographer and was the overall winner of the David Malin Awards for astrophotography in 2005 and 2011.

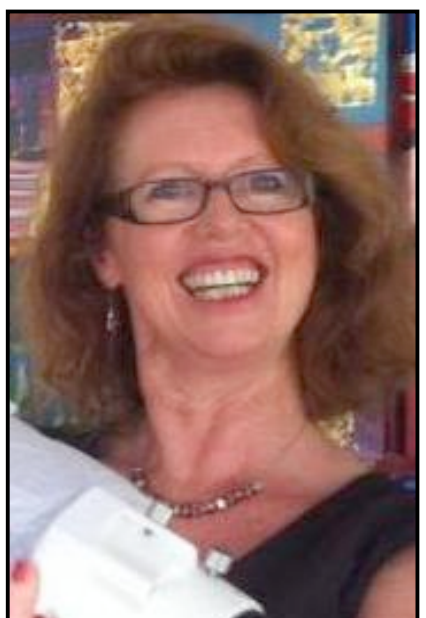

Toner Stevenson is the Manager of Sydney Observatory, part of the Museum of Applied Arts and Sciences. She earned a graduate degree in Design, a Master of Arts in arts management, and is a PhD candidate in museum studies at the University of Sydney, studying the history and heritage of astrographic plates from Sydney Observatory's past research programs.

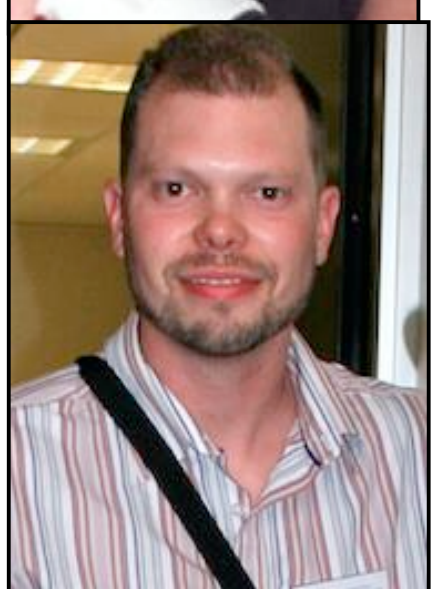

Dr Duane Hamacher is an academic and ARC Discovery Early Career Researcher in the Nura Gili Indigenous Programs Unit at the University of New South Wales in Sydney, Australia. His research and teaching focuses on cultural astronomy, geomythology, and ethnoscience, specialising in Indigenous Australian and Oceanic cultures. He earned graduate degrees in astrophysics and Indigenous studies and works as an astronomy educator and consultant curator at Sydney Observatory.